\documentstyle[preprint,prc,aps,tighten]{revtex}
\begin{document}
\draft
\title{
Nuclear Structure Functions from Constituent Quark Model               
}
\author{Firooz Arash$^{(a,b)}$\footnote{e-mail: arash@Rose.ipm.ac.ir} and
Shahin Atashbar-Tehrani$^{(c)}$}
\address{
$^{(a)}$
Center for Theoretical Physics and Mathematics, Atomic Energy
Organization of Iran, P.O.Box 11365-8486, Tehran, Iran \thanks{Permanent address} \\
$^{(b)}$
Physics Department, Shahid Beheshti University,Tehran, Iran 19834 \\
$^{(c)}$
Physics Department, Amir Kabir University of Technology,
Hafez Avenue, Tehran, Iran \\
}
\date{\today}
\maketitle
\begin{abstract}
We have used the notion of the constituent quark model of nucleon, where
a constituent quark carries its own internal structure, and applied it 
to determine nuclear structure functions ratios. It is found that the description
of experimental data require the inclusion of strong shadowing effect for
$x<0.01$. Using the idea of vector meson dominance model and other ingredients
this effect is calculated in the context of the constituent quark model. It
is rather striking that the constituent quark model, used here, gives a
good account of the data for a wide range of atomic mass number from
$A=4$ to $A=204$.
\end{abstract}

\section {Introduction}

The measurement of the nuclear structure function, $F_2^A(x,Q^2)$, in Deep
Inelastic Scattering (DIS) indicates that the structure function of a bound
nucleon is different from a free nucleon. Very precise data on the proton
structure function from HERA \cite{1}\cite{2}\cite{3}\cite{4} and NMC \cite{5} which cover a wide range
of kinematics, both in $x$ and $Q^{2}$, sets a stringent demand on the
understanding of nucleon structure function in the nuclear environment.
The difference between the parton distributions in the free and the bound
nucleon, that is, $f_{\frac{i}{A}}(x,Q^2)\neq f_{\frac{i}{p}}(x,Q^2)$
is attributed to the nuclear effects. These effects are characterized in the
structure function ratio
\begin{equation}
R=\frac{F_2^A(x,Q^2)}{F_2^D(x,Q^2)}
\end{equation}
There are a host of interpretations that describe the DIS data on the nuclear
targets pioneered by the European Muon Collaboration (EMC effect) and it is
believed that various mechanisms are responsible for the behavior of $R$ in
various kinematical intervals:\\
 i)The Shadowing effect at $x<0.1$ where $R\leq 1.$\\
ii)Anti-shadowing effect at $0.1\leq x\leq 0.3$ where $R\geq 1$.\\
iii)The {\it{EMC}} effect for $0.3\leq x\leq 0.7$ where $R\leq 1$.\\
iv)The Fermi motion for $x>0.7$ where $R\geq 1$.\\
For a detailed study of these mechanisms and other theoretical models
sea Arneode \cite{6}.\\
Currently, there are fairly accurate data on the structure function ratio
$R=\frac{F_2^A(x,Q^2)}{F_2^D(x,Q^2)}$ that has been obtained by different
experimental groups \cite{7} for a variety of nuclei ranging in mass number
from $A=4$ to $A=208$. These data, now, permit us to investigate the $x$ and
$Q^{2}$ dependence of the $F^{A}_{2}(x,Q^2)$. One feature of the data, almost
universal, is the shape of the $x$-dependence of the $F^{A}_{2}$ and is
basically
the same for every nuclei, thus making it rather easy to fit in a wide range
of models with totally different underlying assumptions. Our interest here
is to find a description for the nucleon structure which can be extended
to the nuclei; yet maintaining some universal features common to both. To
this end, our starting point is the study of the internal structure of the
constituent quark which is common to both nucleon and nuclei.
Its internal structure is universal and governed by the annihilation
and pair production in Quantum Chromodynamics. We assert that at high
enough $Q^{2}$, it is the structure of this constituent quark that is
being probed, whereas, at low $Q^{2}$ it behaves as a structure-less
valence quark. Such a model has been applied successfully to the proton
structure for the entire range of kinematics in $x$ and $Q^{2}$ \cite{8}.
Encouraged by this results, we think that the nuclear environment can only
alter the distribution of the constituent quark, leaving its internal
structure intact. The fact that valence quark distribution in nuclei
receives a distortion as opposed to its distribution in a free nucleon
is very similar to the nucleon swelling and the relaxation of the
confinement of the constituent quark within the boundaries of a nucleon
embedded in a nucleus. To put it differently, a nucleon in a nucleus will
relax its boundaries and will occupy broader space, preserving the
universality of the constituent quark of the relaxed nucleon. None of the
number of the constituent quarks, its internal structure, and the average
momentum of the constituent will change. Thus, here we will be dealing
with a collection of structure-full $U$, $D$-type constituent quarks
participating in DIS. 
\section{FORMALISM}
To proceed quantitatively, we need to employ a model which is compatible
with the deep inelastic scattering data on a free nucleon and then we
impose the restrictions due to nuclear effects for the bound nucleons.The
picture that comes in mind is the so-called {\it{Valon}} model of
R.C.Hwa \cite{9}. In
this model the nucleon is considered as a system of bound constituent
quarks, which themselves have structure. The bound state problem is a
nonperturbative effect which is taken into account in the distribution of
constituent quarks. The structure of constituent quark is produced
perturbatively and is free of bound state problem. In such a picture the
structure function of a free nucleon is the convolution of constituent quark
distribution in a nucleon and the structure function of the constituent
quark itself: 
\begin{equation}
F_2^N(x,Q^2)=2e^{2}_{U}\int dy G_{\frac{U}{N}}(y) {\it{f}}^{c}(x,Q^2) +e^{2}_{D}
\int dy G_{\frac{D}{N}}(y) {\it{f}}^{c}(x,Q^2)
\end{equation}
where $G_{\frac {U}{N}}(y)$ and $G_{\frac {D}{N}}(y)$ are the distribution
of $U$ and $D$-type constituents in a
nucleon and ${\it{f_2^C}}(z, Q^2)$ is the structure function of the
constituent calculated in QCD. At sufficiently high $Q^2$, ${\it{f_2^C}}(z,Q^2)$
can be described accurately in the Leading-Order results in QCD. They are also
calculated in the Next-to-Leading Order[8]; but here we will restrict ourselves
to the leading order, because, we are mainly interested in the application of the
model and it describes the nuclear structure functions sufficiently accurately.
The moments of ${\it{f_2^C}}(z,Q^2)$ can be expressed in terms of the evolution
parameter:
\begin{equation}
s=\it{ln}\frac{\it{ln}\frac{Q^{2}}{\Lambda^{2}}}{\it{ln}\frac{Q_{0}^{2}}
{\Lambda^{2}}}
\end{equation}
where $Q_{0} =0.233 {\frac{GeV}{c}}^2$ and $\Lambda=0.345 {\frac{GeV}{c}}^2$ are scale parameters determined from the data.
The moments of the singlet and non-singlet constituent quark structure function
in the leading order solution of the renormalization group equation are given
as[9]:
\begin{equation}
\begin{array}{c}
M^{NS}(n,Q^2)=\exp (-d_{NS}\ s), \\ 
M^S(n,Q^2)=\frac 12(1+\rho )\exp (-d_{+}\ s)+\frac 12(1-\rho )\exp (-d_{-}\
s),
\end{array}
\end{equation}
The anomalous dimensions, $d'^{s}$, and other associated parameters are : 
\begin{equation}
\begin{array}{c}
\rho =(d_{NS}-d_{gg})/\Delta , \\ 
\Delta =d_{+}-d_{-}=[(d_{NS}-d_{gg})^2+4d_{gQ}d_{Qg}]^{1/2}, \\ 
d_{NS}=\frac 1{3\pi b}[1-\frac 2{n(n+1)}+4\sum_{2}^{n}\frac 1j],\\
d_{gQ}=\frac{-2}{3\pi b}\frac{2+n+n^2}{n(n^2-1)},\\
d_{Qg}=\frac{-f}{2\pi b}\frac{2+n+n^2}{n(n+1)(n+2)}, \\ 
d_{gg}=\frac{-3}{\pi b}[-\frac 1{12}+\frac 1{n(n-1)}+\frac 1{(n+1)(n+2)}-\frac
f{18}-\sum _{2}^{n}\frac 1j], \\
d_{\pm }=\frac
12[d_{NS}+d_{gg}\pm \Delta ], \\ 
b=(33-2f)/12\pi ,
\end{array}
\end{equation}
Since these moments, $M(n,s)$, are known, using inverse Mellin transformation
technique, we can obtain the distributions of various components in the
constituent quark; namely, valence, sea quarks, and gluon distributions,
$P^{v(sea, g)}$. They are as follows:
\begin{equation}
\begin{array}{c}
P^{v(s)g}(z,Q^2)=\frac 1{2\pi i}\int_cdnz^{-n+1}M_{v(s)g}(n,s), \\ 
M_v(n,s)=M^{NS}(n,s), \\ 
M_s(n,s)=(2f)^{-1}(M^S(n,s)-M^{NS}(n,s)), \\ 
M_g(n,s)=M_{gQ}(n,s),
\end{array}
\end{equation}
and $M_{gQ}(n,s)$ is the quark-to-gluon evolution function given by: 
\begin{equation}
M_{gQ}(n,S)=\Delta ^{-1}d_{gQ}[\exp (-d_{+}s)-\exp (-d_{-}s)],
\end{equation}
To account for the $SU(2)$ asymmetry of the nucleon sea which is evident
form the violation of the Gottfried sum rule, we follow the same procedure
as in\cite{7} and replace the distribution of d-quark in the sea by: 
\begin{equation}
P_{d/C}(z,Q^2)=\frac 1{(1-x)}P_{u/C}(z,Q^2),
\end{equation}
The calculation of the moments given in eq.(4) is simple. Instead of
exhibiting the moment distribution, we present the results in parametric
form. That is, for every $s$ we fit the moments by a form give below: 
\begin{equation}
\begin{array}{c}
P_{C}^v(z,Q^2)=a_v(Q^2)\ z^{b_v(Q^2)}\ (1-z)^{c_v(Q^2)}, \\ 
P_{C}^{s}(z,Q^2)=\sum_{1}^{2}a_{s_i}(Q^2) (1-z)^{b_{s_i}(Q^2)}, \\
P_{C}^{g}(z,Q^2)=\sum_{1}^{2}a_{g_i}(Q^2) (1-z)^{b_{g_i}(Q^2)}
\end{array}                                                     
\end{equation}
for the valence, sea quarks, and gluon respectively. The dependence of the
coefficients $a_j$, $b_j$ and $c_j$ on $Q^{2}$, or rather on $s$ is given in the
appendix. To complete the calculation of $F_{2}^{N}$, we also need to specify
constituent quark distributions, $G_{\frac{U}{N}}$ and $G_{\frac{D}{N}}$.
The simplest approach to evaluate these distributions is to write down the inclusive
momentum distribution of the constituent quark in a free nucleon. We will limit
ourselves to proton here. these inclusive distributions can be written as:
\begin{equation}
G_{UUD}(y_1,y_2,y_3)=\alpha \ y_1^a\ y_2^a\ y_3^b\ \delta (y_1+y_2+y_3-1)
\end{equation}
where $a=0.65$ and $b=0.35$ are two free parameters and $y_1, y_2$ refer
to $U$-type and $y_3$ refers to the $D$-type constituents. By double
integration
over the unspecified variable we get the exclusive distributions: 
\begin{equation}
\begin{array}{c}
G_{\frac UP}(y)=[B(a+1,a+b+2)]^{-1}\ y^a\ (1-y)^{a+b+1} \\ 
G_{\frac DP}(y)=[B(b+1,2a+2)]^{-1}y^b\ (1-y)^{2a+1}
\end{array}
\end{equation}
where $B(x,y)$ is the Euler Beta function. The above distributions satisfy
also the following normalization conditions: 
\begin{equation}
\int_0^1dy\ G_{\frac UP}(y)=\int_0^1dy\ G_{\frac DP}(y)=1
\end{equation}
In figure 1, we present the result for $F_2^P(x,Q^2)$ for several values
of $x$ and $Q^2$, its agreement with the experimental data in a wide range
of kinematics is rather good.\\
Having determined the nucleon structure function, next we extend the same
procedure to the nuclear medium. As we mentioned earlier, in the nuclear
medium the constituent quark distribution has to be modified,
but its structure function, or rather parton distribution in a constituent
will not be affected by the nuclear environment. The modification to
$G_{\frac CP}(y)$ is achieved by introduction of a distortion factor,
$\delta$. Hence, we write the inclusive momentum distribution as follows : 
\begin{equation}
G_{UUD}^{\prime }(y)=\alpha (y_1y_2)^{a+\delta}y_3^{b+\delta}\delta
(y_1+y_2+y_3-1)
\end{equation}
which leads to : 
\begin{equation}
\begin{array}{c}
G_{\frac UP}^{\prime }(y)=\alpha \ y^{a+\delta}\ (1-y)^{a+b+2\delta+1} \\ 
G_{\frac DP}^{\prime }(y)=\alpha ^{\prime }\ y^{b+\delta}\
(1-y)^{2a+2\delta+1} 
\end{array}
\end{equation}
with : 
\begin{equation}
\begin{array}{c}
\alpha =[Beta(a+\delta+1,a+b+2\delta+2)]^{-1} \\ 
\alpha ^{\prime }=[Beta(b+\delta+1,2a+2\delta+2)]^{-1} 
\end{array}
\end{equation}
we note that $\delta$ is a function of atomic number, $A$. In Ref.\cite{8} a
probabilistic argument is used to find $\delta$. Here we do generalize
the same result as follows : 
\begin{equation}
\delta=\frac{0.0104-0.014 \ln A+ 0.014(\ln A)^2}{%
1-0.83 \ln A+0.295(\ln A)^2-0.024(\ln A)^3}
\end{equation}
So far we have determined the constituent quark distribution in a bound
nucleon. We need to find its distribution in a nucleus. For that matter,
however, one needs also to know the nucleon distribution in the nucleus.
we can use the Fermi distribution and arrive at the following result : 
\begin{equation}
\phi _N^A(z)=\int_{k_F}^{\left| k\right| }d^3k\ \rho (k)\ \delta (z-1-\frac{%
k^{+}}{M_n})
\end{equation}
where $k^{+}=k^0+k^z$ and $k^0=M_{N}$ are the light cone coordinates
and $\rho (k)=\frac 3{4\pi k_F^3}\theta (k_F-\left| k\right| )$ is the nucleon
density in the nucleus. So,
straightforward integration would yield the nucleon distribution in a
nucleus. However, this also means that the bound nucleon is nearly on shell
and therefore, the nuclear binding effect is neglected. To be more
realistic, and include this effect, we replace $z-1$ by $z-\eta _A$ where $%
\eta _A=1-\frac{B_A}{M_N^{*}}$ and  $M_N^{*}=M_N-B_A$ is the effective mass
of the nucleon. With these fine tunings we arrive at the following function
for the nucleon distribution in a nucleus : 
\begin{equation}
\begin{array}{c}
\phi _N^A(z)=\frac 34
\frac{M_N^3}{k_F^3}(\frac{k_F^2}{M_N^2}-(z-\eta _A)^2)\ ;\eta _A-\frac{k_F^A
}{M_N}\langle \ z\ \langle \eta _A+\frac{k_F^A}{M_N} \\ 
\phi _N^A(z)=0\ ;otherwise
\end{array}
\end{equation}
The values of the Fermi momentum, $k_F^A$, and the binding energy, $B_A$,
are given by the requirements of nuclear physics. Going one step further, we
carry the Fermi motion to the constituent quark level. This is done
easily by renormalization of the distribution function give in Eq.(14) this
leads to the following constituent quark distribution per proton in a nucleus : 
\begin{equation}
G_{\frac{U}{P}}^{A}(y)\approx (\frac 1\eta )G_{\frac UP}^{\prime }(\frac y\eta
)+\frac {1}{10}\lambda ^2(
\frac{d^2}{dz^2})(\frac 1z)G_{\frac UP}^{\prime }(\frac yz)\mid _{z=\eta }
\end{equation}
\begin{equation} 
G_{\frac DP}^A(y)\approx (\frac 1\eta )G_{\frac DP}^{\prime }(\frac y\eta
)+\frac 1{10}\lambda ^2(\frac{d^2}{dz^2})(\frac 1z)G_{\frac DP}^{\prime
}(\frac yz)\mid _{z=\eta }
\end{equation}
where $\lambda =\frac{k_F}{M_N^{*}}$ .
Having collected all the ingredients, now we are
in a position to evaluate the nuclear structure function ratios,
$R=\frac{F_2^A(x,Q^2)}{F_2^D(x,Q^2)}$. Notice that all of our discussions about the constituent quark distribution are pertinent to proton. Finally, 
taking the neutron excess in the nucleus into consideration, we get : 
\begin{equation}
F_{2}^{A}(x, Q^2)=(\frac{1}{A})\{F_{2A}(x, Q^2)-\frac{1}{2}(N-Z)[F_2^n(x,Q^2)-F_2^p(x,Q^2)]\}
\end{equation}
where $F_2^p(x,Q^2)$ and $F_2^n(x,Q^2)$ are the free proton and free neutron
structure functions, respectively, and $F_{2A}(x,Q^2)$ is obtained using
Eq.(1). The results of the model is presented in figure $2$ for a variety of
nuclei ranging from $A=4$ to $A=204$. As it is apparent from Fig.(2), we
see that the model calculation is reasonable only down to $x\geq 10^{-1}$.
For smaller values of $x$, $x\leq 10^{-1}$ an extra ingredient is
required. An obvious and natural place to look for the dynamics of small $x$
would be the shadowing effects, where the growth of parton densities are
suppressed due to recombination and annihilation of the partons. There are
some attempts in the literature to calculate the shadowing corrections,
pioneered by Qiu\cite{12}. However we found that such a
parameterization is inadequate in describing the data. Following Ref.[12] we
modify the shadowing correction of Qiu in two ways: (1) we take the
general idea of vector meson dominance which describes low $Q^2$ virtual
photon interaction, to model the approximate scaling of shadowing by pomeron
exchange. It turns out that the experimental data on $F_{2p}(x,Q^2)$ up to $%
x<0.05$ can be well described by a sub-asymptotic pomeron with the effective
anomalous dimension going down logarithmically with $x$\cite{13}.(2) The magnitude of the shadowing
effect in such models is sensitive to the value of vector meson-nucleon
cross section. This sensitivity is explored by G. Show \cite{14} and it is
described by a quantity : 
\begin{equation}
\frac{A_{eff}}A=\frac{\sigma _{\gamma \ A}}{A\sigma _{\gamma \ N}}
\end{equation}
This ration is particularly manifested for heavy nuclei and reflects the
ratio of nuclear radius $R$ to the mean free path, $\lambda$, of the
vector meson inside the nucleus; which also depends on the nuclear density .
That is, the shadowing effect at small $x$ values increases with increasing
nuclear radius and density. Utilizing these modifications to the original
work of Qiu [12], finally we present the following modification of parton
distributions in a nucleus : 
\begin{equation}
\begin{array}{c}
P_{C/A}^s(z,Q^2)=R_s(x,Q^2,A_{eff})P_{C/N}^s(z,Q^2), \\ 
P_{C/A}^v(z,Q^2)=P_{C/N}^v(z,Q^2),
\end{array}
\end{equation}
where superscript $s(v)$ stands for sea (valence) components and 
$R_s(x,Q^2,A_{eff})$ is the shadowing factor affecting parton distributions
of the constituent quark. These distributions in the nuclear medium are
compared to the corresponding distribution in a free nucleon. The
parameterized form of this factor is : 
\begin{equation}
R_{s}(x, Q^2, A_{eff})=R_{s}(Q^{2},A_{eff})(0.94+0.076A_{eff}^{0.5})x^{0.5}ln(x))
\end{equation}
And $A_{eff}$ is given by: 
\begin{equation}
A_{eff}=-10.97-0.704A+.0042A^2-0.00019A^{2.5}+8.63A^{0.5} 
\end{equation}
In equation (24) $R_s(Q^2,A_{eff})$ is evaluated by the $n^{th}$-moment
equations for the parton distribution in the sea of a constituent quark\cite{15}:
\begin{equation}
\begin{array}{c}
<F_{c/N}^s(Q^2)>_n=<F_{c/N}^s(Q_c^2)>_nK_{qq}^n(Q^2)+<F_{c/N}^g(Q_c^2)>_nK_{qg}^n(Q^2)+ \\ 
<F_{c/N}^v(Q_c^2)>_nK_{NS}^n(Q^2)
\end{array}
\end{equation}
where $<F(Q^2)>_n=\int_0^1z^{n-2}P(z,Q^2)dz$. Obviously, nuclear shadowing
do not affect the distribution of valence quarks in the constituent quarks.
If the momentum loss owing to the recombination process of the shadowed sea
quarks is negligible, the momenta carried by each parton component would be
conserved separately and approximately, 
\begin{equation}
\begin{array}{c}
<F_{c/N}^s(Q^2)>_2\cong <F_{c/A}^s(Q^2)>_2 \\ 
<F_{c/N}^g(Q^2)>_2\cong <F_{c/A}^g(Q^2)>_2
\end{array}
\end{equation}
and the following inequalities will be satisfied for $Q^2> Q_{0}^{2}$[15] : 
\begin{equation}
<F_{c/N(A)}^{s(g)}(Q^2)>_2\gg <F_{c/N(A)}^{s(g)}(Q^2)>_3 
\end{equation}
As a result, following Re.\cite{12} we acquire:
\begin{equation}
R_sea(Q_0^2,A_{eff})=1-k_s(Q_0^2)(A_{eff}^{1/3}-1)=\lim_{x\rightarrow 0}
\frac{F_A^s(x,Q^2)}{F_N^s(x,Q_0^2)}\approx \frac{<F_{C/N}^s(Q_0^2)>_3}{%
<F_{C/A}^s(Q_0^2)>_3} 
\end{equation}
\begin{equation}
R_g(Q_0^2,A_{eff})=1-k_g(Q_0^2)(A_{eff}^{1/3}-1)=\lim_{x\rightarrow 0} 
\frac{F_A^g(x,Q_0^2)}{F_N^g(x,Q_0^2)}\approx \frac{<F_{C/N}^g(Q_0^2)>_3}{%
<F_{C/A}^g(Q_0^2)>_3} 
\end{equation}
where $K_{s(q)}$ is the shadowing strength and and depends on the starting scale
$Q_{0}^{2}$ for the evolution to set in.
It is easy to verify that the primitive shadowing for gluon in a constituent
quark is weaker than that for the sea partons[12] at such scale.The form of 
shadowing correction is in terms of evolution kernels is given as:
\begin{equation}
\begin{array}{c}
R_s(Q^2,A_{eff})= 
R_s(Q_0^2,A_{eff})\frac{%
<F_{c/N}^s(Q_0^2)>_3K_{qq}^3(Q^2)+<F_{C/N}^g(Q_0^2)>_3K_{qg}^3(Q^2)+<F_{C/N}^v(Q_0^2)>_3K_{NS}^3(Q^2) 
}{<F_{C/N}^s(Q_0^2)>_3K_{qq}^3(Q^2)+\lambda
<F_{C/N}^g(Q_0^2)>_3K_{qg}^3(Q^2)+R_s(Q_0^2)<F_{C/N}^v(Q_0^2)>_3K_{NS}^3(Q^2)%
} 
\end{array}
\end{equation}
where $K_{ab}^{n}$ are the evolution kernels. Defining $s=\it{ln}L$, these
kernel functions are[15]:
\begin{equation}
\begin{array}{c}
K_{qq}^n(Q^2)=\alpha _nL^{-a_n^{-}}+(1-\alpha _n)L^{-a_n^{+}} \\ 
K_{qg}^n=\beta _n(L^{-a_n^{-}}-L^{-a_n^{+}}) \\ 
K_{NS}^n(Q^2)=\left[ \alpha _nL^{-a_n^{-}}+(1-\alpha
_n)L^{-a_n^{+}}-L^{-a_{NS}^n}\right] \\ 
a_{NS}= 
\frac{\gamma _{qq}}{\beta _{\circ }}\ ,a_{\pm }^n=\frac{\gamma ^{\pm n}}{%
\beta _{\circ }} \\ \alpha _n= 
\frac{\gamma _{qq}-\gamma ^{+}}{\gamma ^{-}-\gamma ^{+}} \\ \beta _n=\frac{%
\gamma _{qq}}{\gamma ^{-}-\gamma ^{+}} 
\end{array}
\end{equation}
\section {CONCLUSION}
We have calculated the nucleus structure function ratio in the context of
constituent quark picture, utilizing the essence of the so called valon model.
As we can see from the figures; the model describes the data rather well for
a wide range of $A$. It appears that the $Q^2$ dependence of the ratio of
structure functions is week. The shadowing effect is quite large for small
value of $x$, $(x\langle 0.01)$ and dies away very rapidly as $x$ increases.
Also, the shadowing correction becomes more pronounced with increasing
atomic number $A$. We further note that in this model, which is based on
the constituent quark structure, we have not included explicitly the
anti-shadowing effects notwithstanding that there are some debates on the
subject particularly pertinent to the middle range of $x$. we simply did
not need it to describe the data. Our anticipation is that such an effect
would not be very large .
\section{APPENDIX}
In this appendix we provide numerical values for the coefficients given in
equation ( ). These are the results of our $Q^2$ dependence parameterization
of parton distributions in a constituent quark of proton. They are calculated
in the leading order. 
\begin{equation}
\begin{array}{c}
a_{s1}=-0.119+0.0296\exp (s/0.66) \\ 
b_{s1}=5.510+8.50\exp (s/0.95) \\ 
a_{s2}=-0.080+0.066\exp (s/3.12) \\ 
b_{s2}=-1.194+2.303\exp (s/2.76) \\ 
a_{g1}=-2.490+1.027\exp (s/0.69) \\ 
b_{g1}=5.920+6.440\exp (s/0.80) \\ 
a_{g2}=.174+.345s^{0.09} \\ 
b_{g2}=-1.407+1.763\exp [s/2.42]; \\ 
a_v=-0.300+1.254s-0.368s^2+0.034s^3 \\ 
b_v=0.196+1.286\exp [-s/3.14] \\ 
c_v=-0.844+.451s^{.17} 
\end{array}
\end{equation}
\newpage
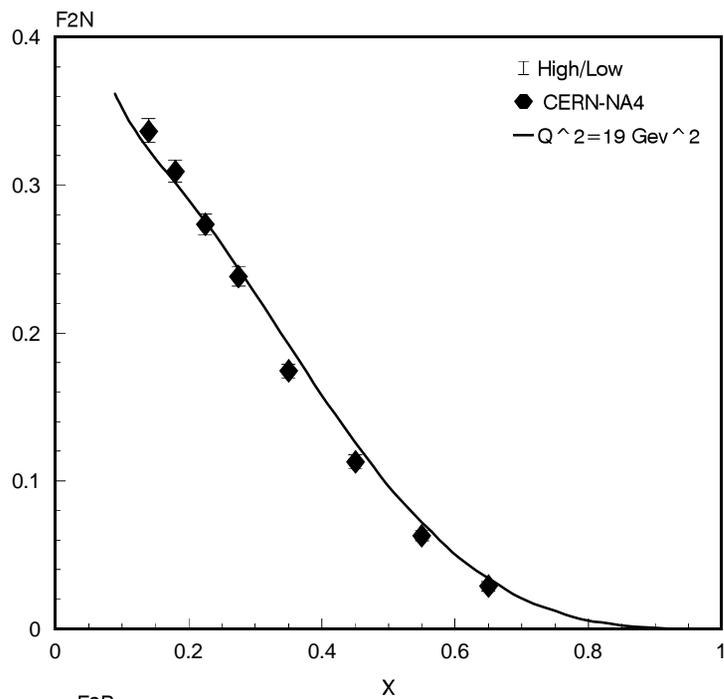
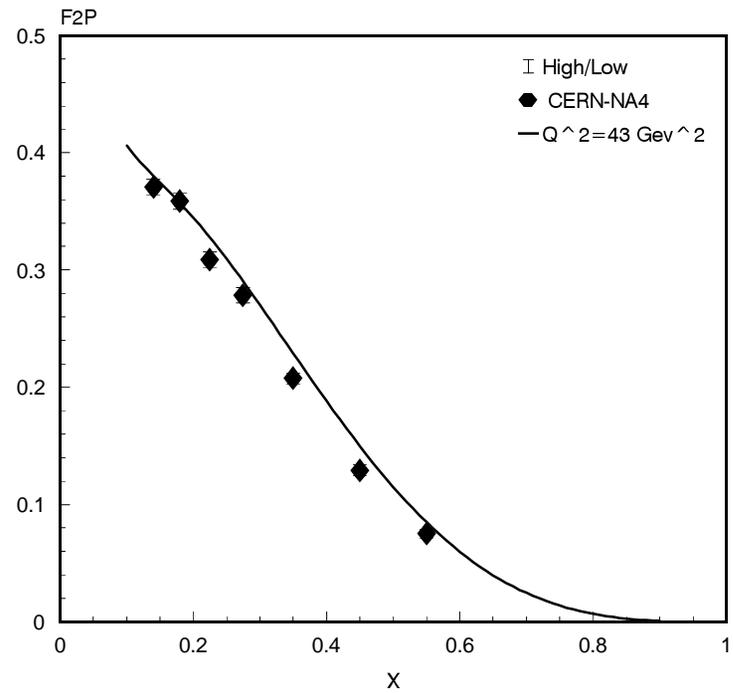
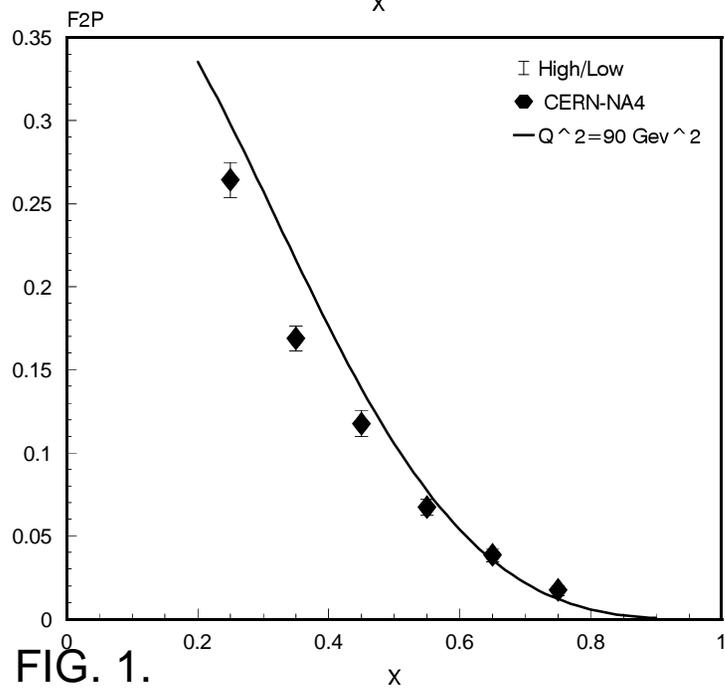
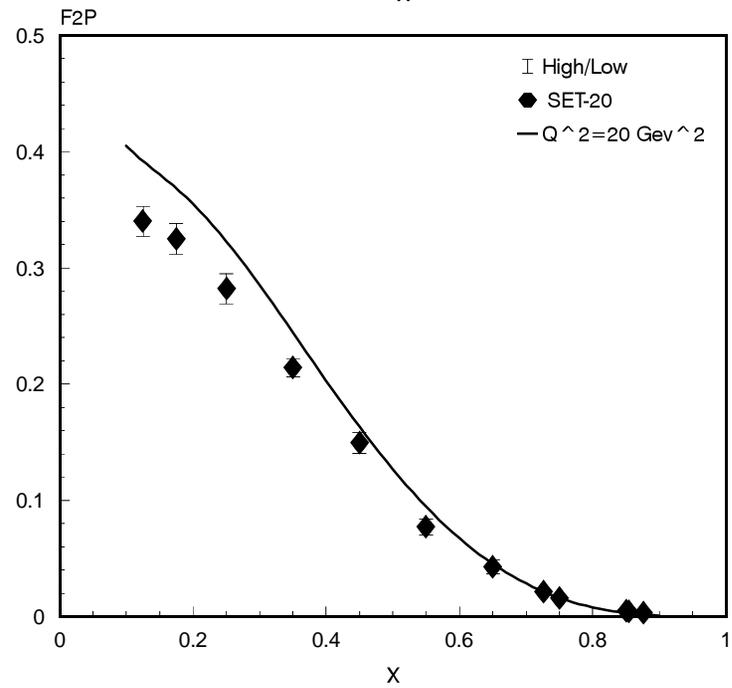
\begin{figure}
\label{fig1}
\caption{ $F^{p}_{2}$ as a function of $x$ at several $Q^{2}$.}  
\end{figure}
\begin{figure}
\label{fig2}
\caption{ The ratio $R=\frac{F_2^{A_1}(x,Q^2)}{F_2^{A_2}(x,Q^2)}$ for
a variety of nuclei at different $Q^2$. Dashed-dotted line represents
the results without the shadowing effects. Others include the shadowing
corrections.}
\end{figure}


\newpage
\begin{references}
\bibitem{1}H1 Coll., C. Adloff, {\it{et al.}}, Nucl. Phys. B {\bf{497}}, 
3 (1997). ZEUS Coll., J. Breitweg, {\it{et al.}}, DESY-98-121 (1998). 
ZEUS Coll., M. Adamus, {\it{et al.}}, Phys Lett. B {\bf{407}} (1997) 432.
ZEUS Coll., M. Derrick, {\it{et al.}}, Z. Phys. C {\bf{72}} (1996) 399.
\bibitem{2}M. Arneodo,{\it{et al.}}, Nucl. Phys. B {\bf{333}} (1990)1;
P.Amaudruz, {\it{et al.}}, Phys. Lett. B {\bf{295}} (1992) 159.
\bibitem{3}H1 Collaboration, T.Ahmed,{\it{et al.}},Nucl. Phys. B {\bf{439}} 
471 (1995),{\it{ibid}}, Nucl.Phys. B {\bf{407}} (1993) 515.
\bibitem{4}ZEUS Collaboration. M. Derrick, {\it{et al.}} Z. Phys. C {\bf{65}},
R.G.Roberts and M.R. Whalley, J. Phys. G{\bf{17}}, (1991) D1. 293 (1995).
\bibitem{5}NMC Collaboration, P.Amaudruz {\it{et al.}}
Phys.Rev.Lett. {\bf{66}} (1991) 2712;Phys.Rev. D {\bf{50}} (1994) R1.
\bibitem{6} M. Arneodo, Phys. Rep.{\bf{240}},301 (1994).
\bibitem{7}NMC Coll., M. Arneodo, {\it{et al.}}, Nucl. Phys. {\bf B481} (1996) 3. P. Amaudruz, {\it{et al.}}, Nucl. Phys {\bf {B441}} (1995) 3.
\bibitem{8} F. Arash and A. N. Khorramian, {\it{Next-to-Leading Order calculation
of the Nucleon Structure Function In the Constituent Quark Model}}, submitted to
{\it{Phys. Rev. D}}.
\bibitem{9}R.C. Hwa, Phys.Rev. D{\bf{22}} (1980) 1593. R.C. Hwa, Phys.
Rev. D{\bf{51}} (1995) 85.
\bibitem{10}F. Arash and L.Tomio, Phys. Lett. B {\bf{401}} (1997) 207.
\bibitem{11}W. Zhu and L.Qian, Phys. Rev. C{\bf{45}}, 1397, (1992); Phy. Rev. D{\bf{44}}, 2762, (191).
\bibitem{12}Jian-Wei Qiu, Nucl. Phys.{\bf{B291}}, 746, (1987).   
\bibitem{13}N. Armesto and M.A.Branin, Z. Phys. C76 (1997) 81.
\bibitem{14}Graham Shaw, Phys. Rev. D47, R3676 (1993).
\bibitem{15}W. Zhu and J. G. Shin, Phys. Rev. C{\bf{41}}, 1674, (1990).
\end{references}
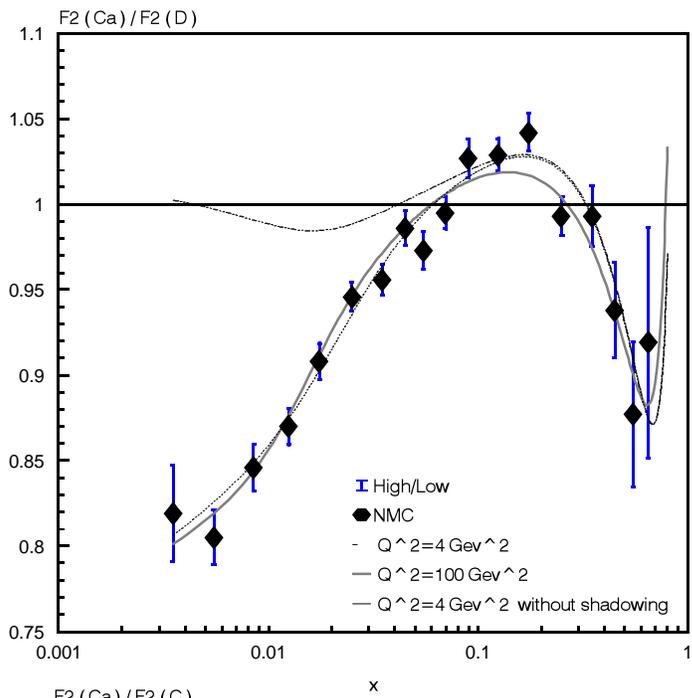
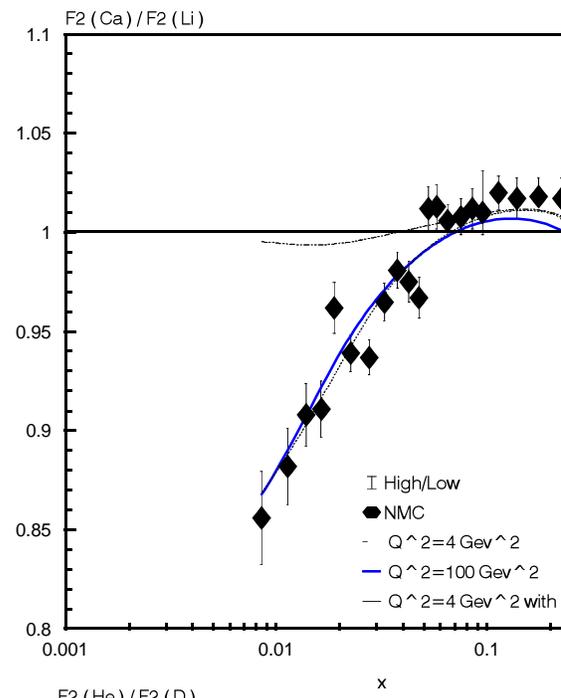
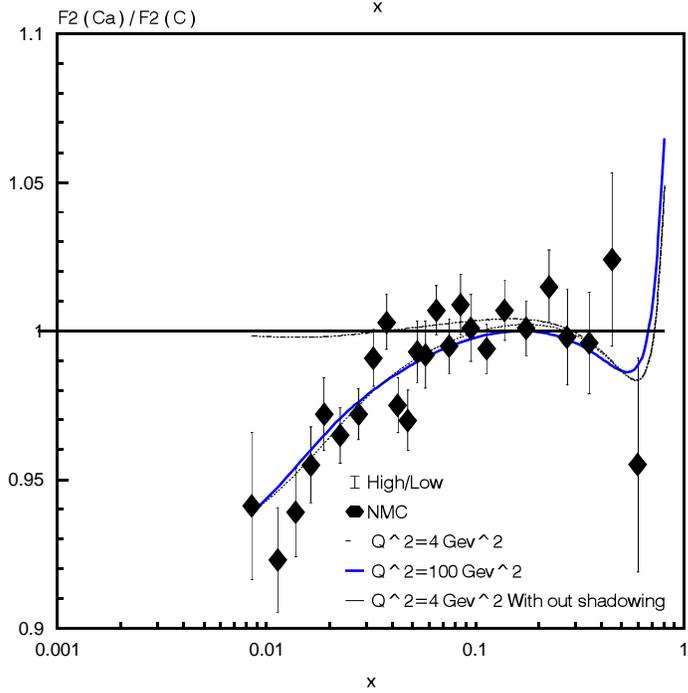
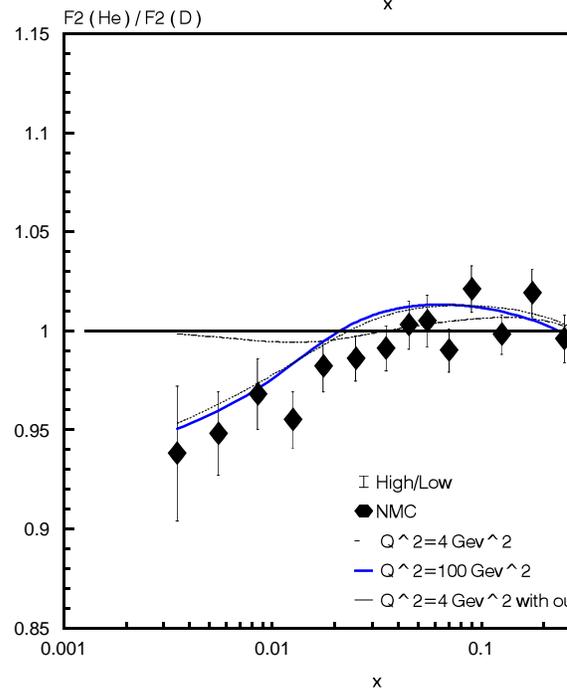
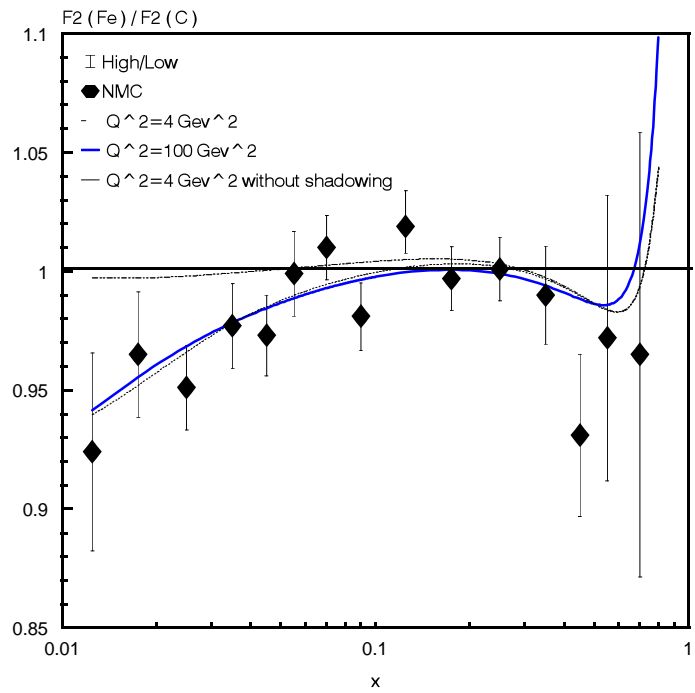
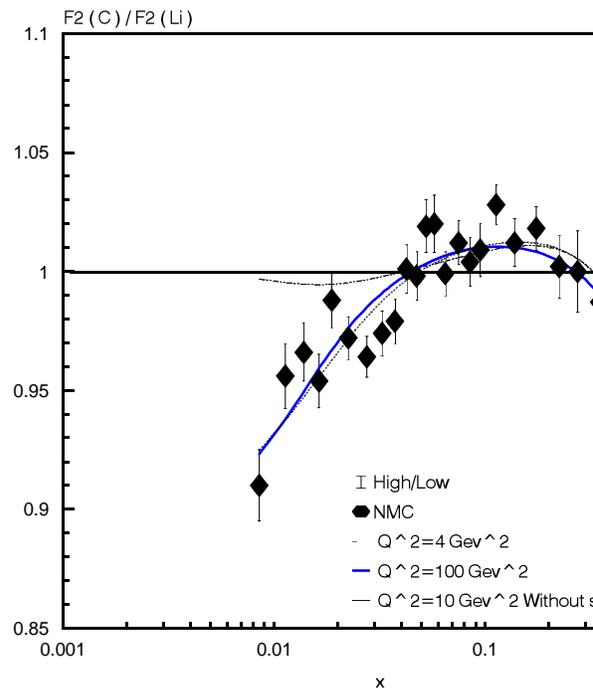
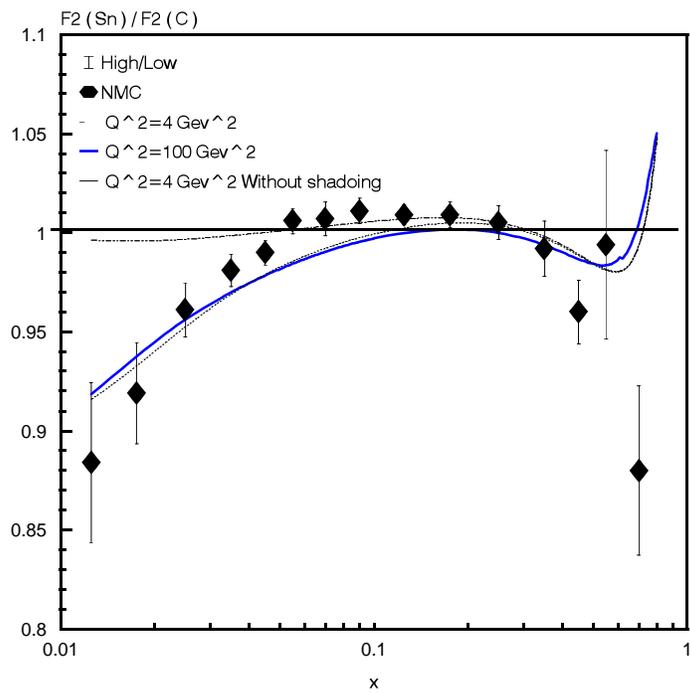
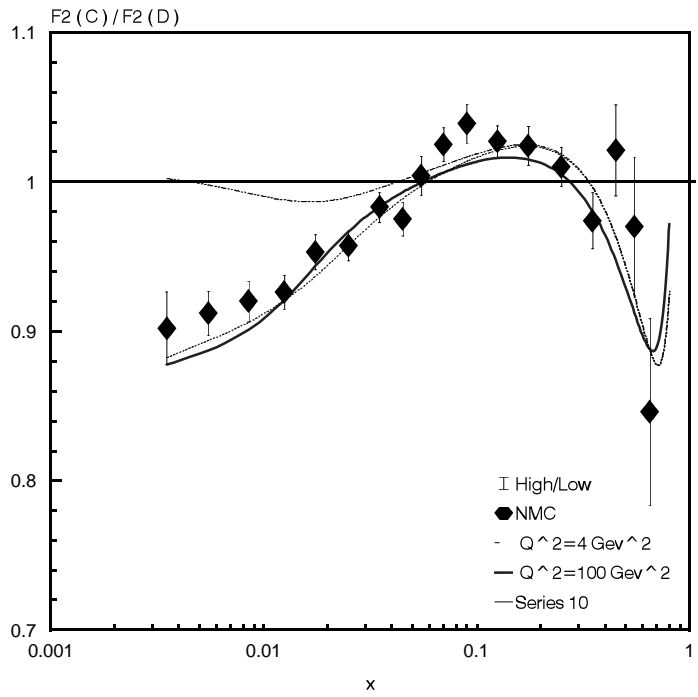
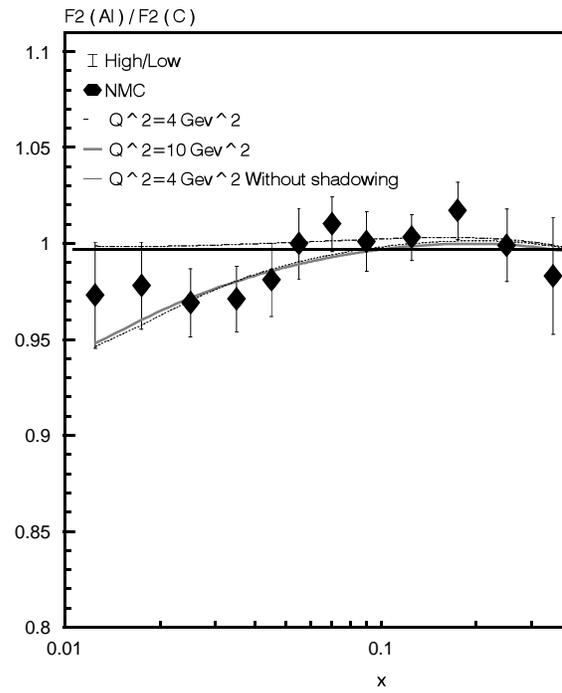
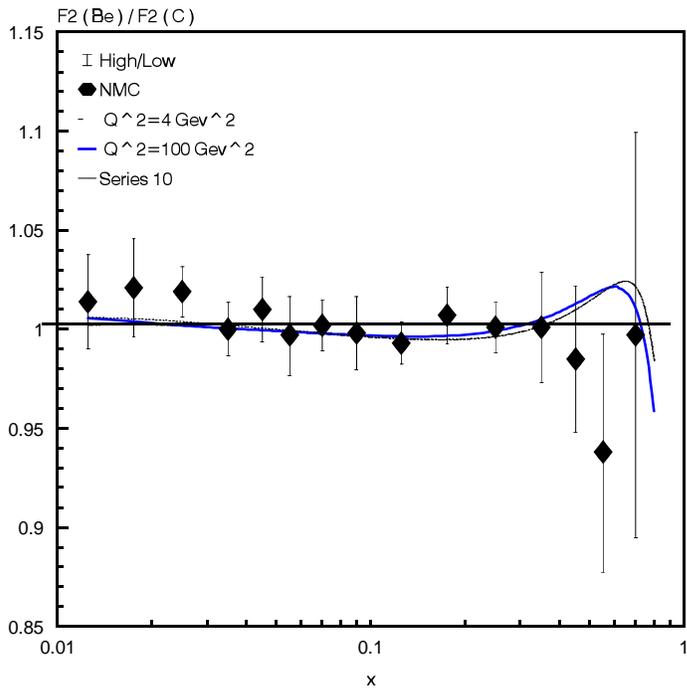
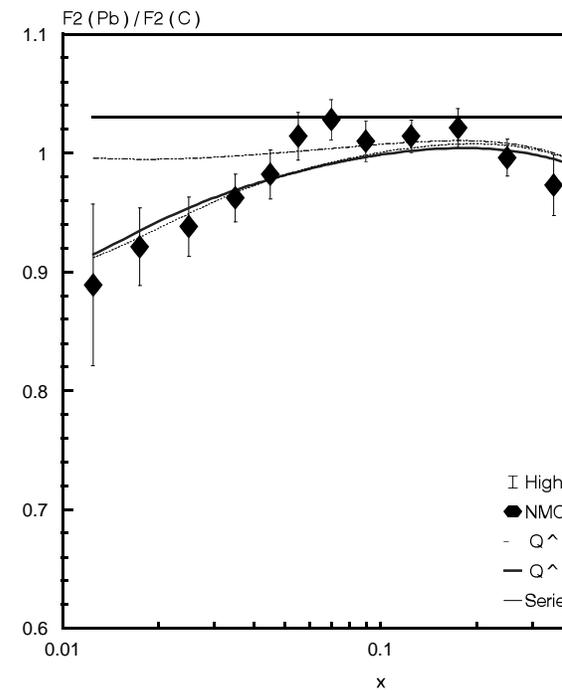
\end{document}